\documentclass{article}
\usepackage{cite,graphicx}
\def\roughly#1{\mathrel{\raise.3ex\hbox{$#1$\kern-.75em\lower1ex%
\hbox{$\sim$}}}}

\def\gtrsim{\roughly>}
\newcommand{\art}[5]{\textit{#1}~\hbox{$//$} #2, #3. V.~#4. P.~#5.}
\newcommand{\book}[3]{\textit{#1} #2, #3.}
\newcommand{\proc}[6]{\textit{#1}~\hbox{$//$} #2. #4. #5. V.~#3. P.~#6.}
\def\PRA{Phys.\ Rev.~A}

\def\PRL{Phys.\ Rev.\ Lett.}
\def\PD{Physica D}

\def\PLA{Phys.\ Lett.~A}

\makeatletter \renewcommand\@biblabel[1]{#1.} \makeatother
\begin{document}
\twocolumn[\hsize\textwidth
          \columnwidth\hsize
          \csname @twocolumnfalse\endcsname
\title{%
     {\normalsize
     \textit{UDC 537.591.15}\hfill\texttt{astro-ph/0205260}}\\[2mm]
     \textbf{NONLINEAR ANALYSIS OF EAS CLUSTERS}
     }
\author{%
     \textbf{M. Yu.~Zotov, G. V. Kulikov, Yu.~A. Fomin}\\[2mm]
     \textit{D. V. Skobeltsyn Institute of Nuclear Physics}\\
     \textit{M. V. Lomonosov Moscow State University, Moscow 119992,
     Russia}
     }
\date{}
\maketitle
\begin{quotation}
\noindent
     We apply certain methods of nonlinear time series analysis
     to the extensive air shower clusters found earlier in the data
     set obtained with the EAS--1000 Prototype array.
     In particular, we use the Grassberger--Procaccia algorithm
     to compute the correlation dimension of samples in the vicinity of
     the clusters.
     The validity of the results is checked by surrogate data tests and
     by some additional quantities.
     We compare our conclusions with the results of similar
     investigations performed  by the EAS-TOP and LAAS groups.
\end{quotation}
\bigskip
]

\section{Introduction}

     We have already studied the distribution of arrival times of
     extensive air showers (EAS) registered with the EAS--1000 Prototype
     array both by methods of classical statistics \cite{Dubna,Hamburg}
     and by methods of cluster analysis \cite{IzvRAN01,clusters'02}.
     In particular, we have found EAS clusters---groups of consecutive
     showers that were registered in time intervals much shorter than
     expected ones \cite{IzvRAN01,clusters'02}.
     Besides this, we have found that as a rule samples which contain
     clusters do not allow one to accept a hypothesis that EAS arrival
     times have an exponential distribution.
     To the contrary, the vast majority of other sufficiently
     long samples satisfy the same hypothesis if the barometric effect is
     taken into account \cite{Dubna,Hamburg}.
     Thus we decided to apply methods of nonlinear time series analysis
     to samples that contain EAS clusters in order to clarify dynamical
     reasons of this situation.
     Below we present some results of this investigation.

     Recall that modern methods of nonlinear time series analysis are
     mainly based on the results obtained by Takens~\cite{Takens80},
     Ma\~n\'e~\cite{Mane}, and Packard et~al.~\cite{Packard-etal}, and
     an algorithm suggested by Grassberger and Procaccia~\cite{GP83}
     and modified by Theiler~\cite{Theiler:W}.
     We are not going to review these results here (see, e.g.,
     \cite{Moon,Schuster,MP:URSS,Schreiber:PhysRep}) but will remind only
     the basic ideas of this approach.
     Suppose one studies a scalar time series $\{x_i\}_{i=1}^n$ which,
     for instance, presents experimental results obtained during
     observations of a (nonlinear) process.
     It was suggested to study delay vectors
     $X_i = (x_i, x_{i+\tau}, x_{i+2\tau},\dots,x_{i+(m-1)\tau})$
     in order to figure out whether the process possesses chaotic
     dynamics or not; here~$\tau$ is an arbitrary but fixed parameter,
     and $m$ is an integer constant called an embedding dimension.
     One should compute the number~$K$ of vectors with mutual
     distance $\le\rho$ and such that delay vectors~$X_i$ are
     shifted by at least~$W$ indices.
     After this, one calculates the correlation dimension
\[
     D_2(\rho) = \frac{d \log C_2(\rho)}{d \log \rho},
\]
     where $C_2 = K/(\hbox{total number of vectors } X_i)$ is the
     correlation sum.
     If the plot of~$D_2$ has a plateau, then it is likely that the
     process demonstrates chaotic dynamics.
     A value of~$D_2$ at the plateau is taken as an estimate of the
     correlation dimension of the attractor underlying the data.
     This quantity also gives a (lower) estimate for the number of
     degrees of freedom in the process under consideration.

     Investigations in this field have lead to a discovery that a plateau
     in the plot of the correlation dimension can be observed not only
     for chaotic deterministic processes but also for certain types of
     stochastic processes \cite{OsbPro,PSVM}.
     Thus there appeared a problem to distinguish these two classes using
     experimental data.
     The problem has occurred to be complicated, and this called an
     appearance of a number of approaches to its solution.
     One of the main approaches is the method of surrogate data
     \cite{Theiler-etal92,TheilerPrichard96}.
     Other methods are based on certain additional functions.
     As for surrogate data, they are used both to compute some quantities
     that allow one to estimate a measure of nonlinearity of the process
     under consideration and to clarify the ``nature'' of the plateau in
     the plot of~$D_2$.
     Namely, if the plateau is observed for surrogate data, then the
     process is assessed to be stochastic.
     Conversely, if the plateau disappears, then the process is
     deterministic.
     At the moment, there is a variety of methods for generating
     surrogate data.
     One can find a review of the main approaches
     in~\cite{Schreiber:surrogates}.
     Still, we should note that to the best of the authors knowledge
     there is no test that could automatically and unequivocally
     distinguish between a stochastic process and deterministic chaos,
     see, e.g., \cite{MP:URSS,Schreiber:PhysRep}.

\section{The Main Results}

     At the first stage of this investigation the experimental data set
     (time intervals between arrival times of $1.7\times10^6$ EAS
     registered in the period from August, 1997 to February, 1999) was
     split into adjacent samples consisting of 128, 256, and~512
     elements.
     For these samples, we computed the Fourier power spectrum, the
     correlation dimension~$D_2$ (for $\tau=1$, $W=1$, $m=5\dots12$)
     and a number of additional quantities.
     To provide stationarity of the time series under consideration,
     time intervals between EAS arrival times were adjusted to a joint
     value of atmospheric pressure equal to 742~mm~Hg.
     To perform calculations, we used GNU Octave~\cite{Octave}
     running in Mandrake Linux.
     As one could expect, this analysis has revealed that there is no
     plateau in the plot of~$D_2$ for the vast majority of samples.
     Still, in the vicinity of some EAS clusters we have found samples
     that demonstrate signs of chaotic dynamics.  Below we shall briefly
     discuss one of these events, namely a cluster registered on
     November~11, 1998 between 01:21:17.47 and 01:38:02.27 of Moscow
     local time \cite{IzvRAN01,clusters'02}.
     This event consists of three clusters that begin at EAS No.~435,
     436, and~437 respectively and end at EAS No.~570.
     In our opinion, an appearance of these three clusters within one
     event is an effect of our selection procedure~\cite{clusters'02} and
     does not have an astrophysical nature.
     Thus in what follows we shall only discuss the outer cluster, which
     consists of 136~EAS.

\begin{figure}[!t]
\begin{center}
   \includegraphics[width=\hsize]{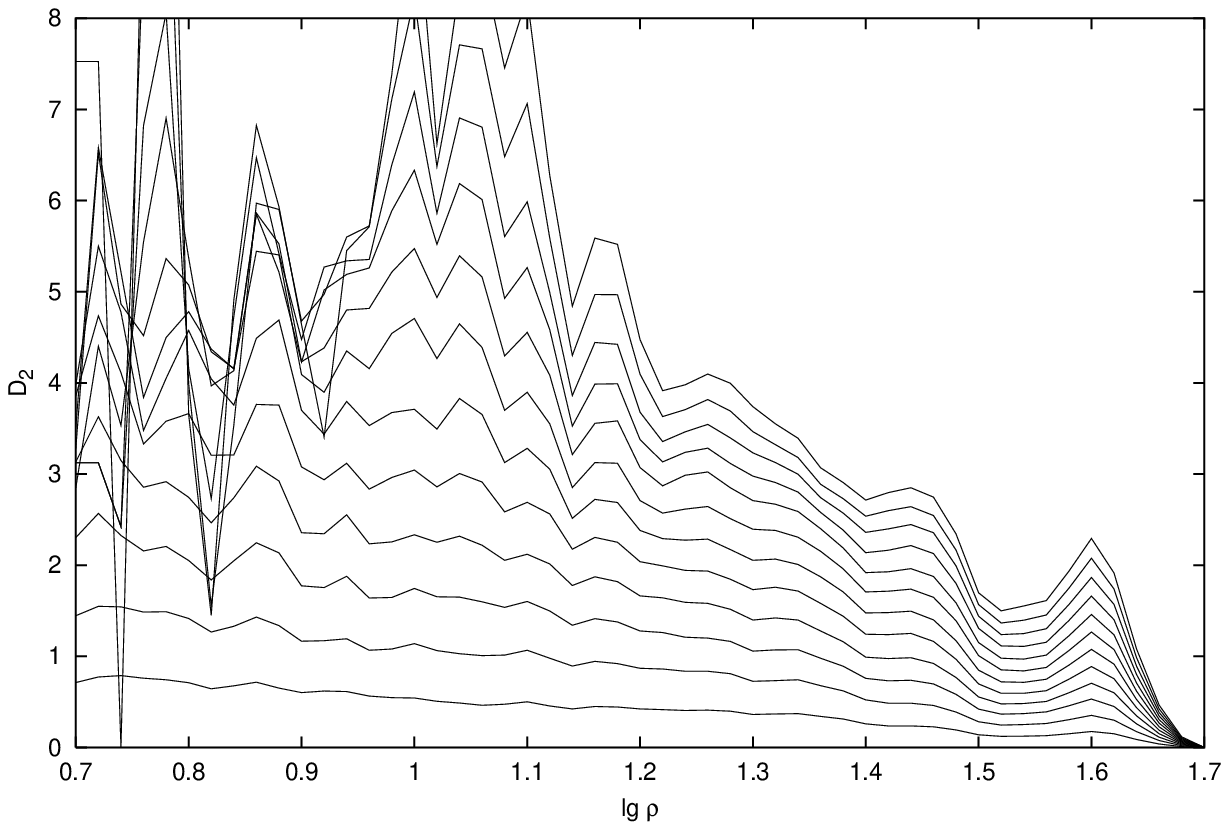}\\
   \includegraphics[width=\hsize]{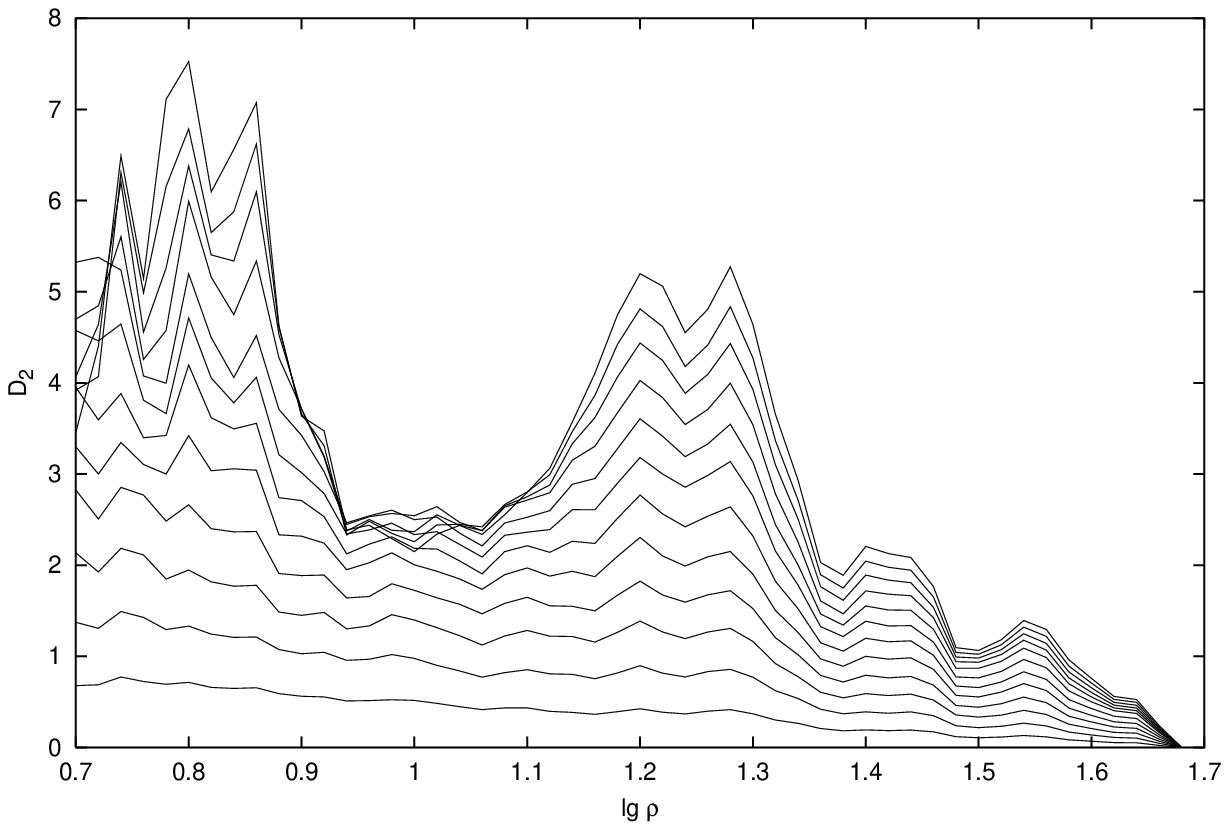}\\
   \includegraphics[width=\hsize]{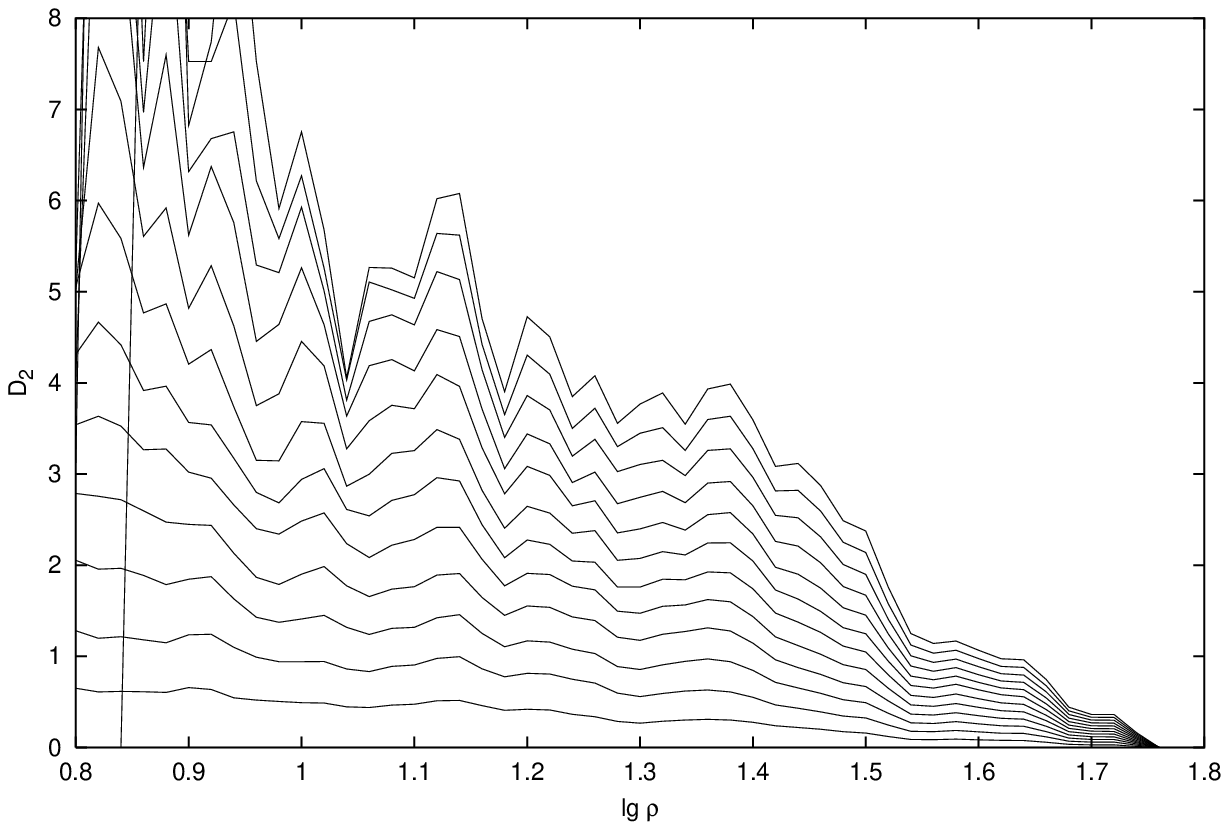}\\
\end{center}
\caption{%
     The correlation dimension $D_2$ for samples that consist of EAS
     No.~168--424, 425--681, and 682--938 (from top to bottom).
}
\end{figure}

     In order to get an idea about the behavior of the correlation
     dimension in the vicinity of the cluster, let us consider three
     adjacent samples, each consisting of 256 time intervals.
     Fig.~1 depicts the correlation dimension for a sample that ends up
     in 2.5 minutes before the cluster, a sample that contains the
     cluster, and a sample that begins in 23 minutes after the cluster.
     The curves were computed for $\tau=1$, $W=1$, and $m=1\dots12$.
     The upper curves correspond to larger values of the embedding
     dimension~$m$.
     The maximum norm was used to compute mutual distances between delay
     vectors.

     As one can see from the figure, the plot of the correlation
     dimension obtained for the sample that contains the cluster has a
     clear plateau with $D_2\approx2.5$.
     On the contrary, no plateau is observed for two other
     samples.
     Thus we conclude that the sample with the cluster demonstrates signs
     of chaotic dynamics with the (fractal) dimension of an attractor
     approximately equal to~2.5.
     At the same time, the Fourier power spectrum of this sample does not
     considerably differ from a broadband spectrum, which can be
     observed, e.g., for random noises.
     We have employed the surrogate data method in order to figure out
     whether the sample with the cluster represents a deterministic
     chaotic process or a stochastic process.
     To make surrogate data, we used two different approaches: a random
     shuffling time delays that constitute the sample, and the
     amplitude adjusted Fourier transform method suggested
     in~\cite{Theiler-etal92}.
     Both methods preserve the distribution of the original data set.
     Besides this, the second method preserves the Fourier power
     spectrum.
     We used the TISEAN package~\cite{TISEAN} to prepare Fourier-based
     surrogate data.

     An analysis of the surrogate data made by both methods for the
     sample that contains the cluster has revealed that plots of the
     correlation dimension do not contain a plateau.
     This gives an argument in favor of the hypothesis for the
     deterministic nature of the original sample since the order of time
     delays that constitute the sample occurs to be important for an
     appearance of the plateau.
     A similar conclusion can be made on the basis of an analysis of
     certain other quantities, e.g., a maximum likelihood estimator for
     the correlation dimension introduced in \cite{Takens85,Theiler88}
     and a function suggested in~\cite{BDSL}.

     On the other hand, we must note that some other tests has lead to
     opposite results.
     For instance, one of the tests for nonlinearity is based on a
     measure for the time-reversibility of a time series~\cite{SS99}.
     An application of this test to the sample that contains the cluster
     and to the surrogates has revealed that the null hypothesis for the
     linear structure of the time series cannot be rejected.
     A possibly stochastic nature of the observed behavior of the
     correlation dimension was also revealed by the analysis of the
     normalized slope, introduced in~\cite{PK98}.

     It is worth mentioning that a plateau in the plot of~$D_2$ for
     samples that partially or completely include the cluster
     can be observed in a wide range of sample lengths (from $N\sim100$
     up to $N\sim500$) and values of the Theiler window~$W$.
     The value of the correlation dimension varies depending on~$N$
     and~$W$ and the position of the cluster inside a sample.
     For example, for the same sample with $N=256$, $D_2\approx3$ for
     $W\ge7$.


\section{Discussion}

     The results presented above demonstrate that one can observe an
     unusual dynamics of EAS arrival times in the vicinity of certain
     clusters of EAS with the electron number of the order of~$10^5$.
     Still it is rather difficult to make a final conclusion on the
     nature of this phenomenon: Does it represent deterministic chaos or
     a special type of a stochastic process?
     In our opinion, the majority of the tests performed witness in favor
     of the first of these two alternatives.
     On the other hand, it is not easy to suggest an astrophysical model
     that could explain chaotic dynamics in EAS arrival times.
     Thus it is interesting to compare our results with the conclusions
     of similar investigations performed by other research groups.

     In a considerable number of articles devoted to the nonlinear
     time series analysis, one can find a comprehensive investigation of
     EAS arrival times registered with the EAS-TOP array~\cite{Aglietta}.
     Basing on a detailed study of the available experimental data set
     and the results obtained with the underground muon
     monitor~\cite{Bergamasco92} the authors of this work made a
     conclusion that though an existence of deterministic chaotic effects
     in cosmic ray time series cannot be completely excluded, cosmic ray
     signals are all color random noise, independently of the nature of
     the secondary particle and of the primary parent particle.
     It was also demonstrated in one of the following articles that
     an impact of background noise brings additional difficulties
     to the problem of distinguishing between chaotic and stochastic
     dynamics~\cite{Bergamasco94}.

     Besides this, a whole series of investigations devoted to the
     nonlinear analysis of EAS time series is carried out in Japan
     beginning from early nineties at the experimental arrays that now
     constitute the LAAS network, see~\cite{Japan2001} and references
     therein.
     The authors of these investigations presented several dozens of
     events that demonstrate chaotic dynamics.
     More than this, it was conjectured that the observed dynamics
     may be due not only to the chaotic structure of the medium
     through which particles have traversed but also to the nature of the
     primary particles~\cite{Japan:particles}.
     Later on, there was suggested a model according to which chaotic
     events may be generated by cosmic rays that have a structure of a
     fractal wave arriving from a nonlinear accelerator like a supernova
     remnant~\cite{Japan:FractalWave}.
     This model needs to be studied in details, but seems to be
     promising.

     Thus, the results obtained during our analysis do not contradict the
     conclusions of similar investigations performed at other EAS arrays.
     It seems to be necessary to continue the work in this area and
     to involve some other methods of nonlinear time series analysis.

\medskip
\centerline{$\star\star\star$}
\smallskip

     We gratefully acknowledge numerous useful discussions with
     A.~V.~Igoshin, A.~V.~Shirokov, and V.~P.~Sulakov who have helped
     us a lot with the experimental data set.
     This work was done with financial support of the Federal
     Scientific-Technical Program ``Research and design in
     the most important directions of science and techniques"
     for 2002--2006, contract No.\ 40.014.1.1.1110,
     and by Russian Foundation for Basic Research grant No.\
     02-02-16081.

     Only free, open source software was used for this investigation.



\newpage
\begin{thebibliography}{99}
\frenchspacing
\bibitem{Dubna}
     \art{Vedeneev O.V., Zotov M.Yu. et al.}{Izv. Ros. Akad. Nauk, Ser.
     Fiz.}{2001}{65}{1224}

\bibitem{Hamburg}
     \proc{Fomin Yu.A., Kalmykov N.N. et al.}
     {Proc. 27th ICRC}{1}{Hamburg}{2001}{195}

\bibitem{IzvRAN01}
     \art{Vedeneev O.V., Zotov M.Yu. et al.}{Izv. Ros. Akad. Nauk, Ser.
     Fiz.}{2001}{65}{1674}

\bibitem{clusters'02}
     \textit{Fomin Yu.A., Kulikov G.V., Zotov~M.Yu.}\\
     Clusters of EAS with Electron Number \hbox{$\gtrsim 10^4$}.
     Preprint 2002-9/693, Skobeltsyn Institute of Nuclear Physics, 2002;
     \texttt{astro-ph/0203478}.

\bibitem{Takens80}
     \proc{Takens F.}{Dynamical Systems and Turbulence,
     Warwick, 1980, D.~A.~Rand and L.-S.~Young, eds., Lecture Notes in
     Mathematics}{898}{Berlin, Springer-Verlag}{1981}{366}

\bibitem{Mane}
     \proc{Ma\~n\'e R.}{Dynamical Systems and
     Turbulence, Warwick, 1980, D.~A.~Rand and L.-S.~Young, eds.,
     Lecture Notes in Mathematics}{898}{Berlin, Springer-Verlag}
     {1981}{230}

\bibitem{Packard-etal}
     \art{Packard N.H., Crutchfield J.P. et al.}
     {\PRL}{1980}{45}{712}

\bibitem{GP83}
     \art{Grassberger P., Procaccia I.}{\PRL}{1983}{58}{2387}\hfill\break
     \art{Grassberger P., Procaccia I.}{\PD}{1983}{9}{189}

\bibitem{Theiler:W}
    \art{Theiler J.}{\PRA}{1986}{34}{2427}

\bibitem{Moon}
     \book{Moon F.C.}{Chaotic Vibrations: An Introduction for
     Applied Scientists and Engineers}{Wiley, New York, 1987}

\bibitem{Schuster}
     \book{Schuster H.-G.}{Deterministic Chaos: An Introduction}%
     {Physik Verlag, Weinheim, 1988}

\bibitem{MP:URSS}
     \book{Malinetskii G.G., Potapov A.B.}{Modern Problems of
     Nonlinear Dynamics}{Moscow, Editorial URSS, 2000 (in Russian)}

\bibitem{Schreiber:PhysRep}
     \art{Schreiber T.}{Phys. Rep.}{1999}{308}{1}

\bibitem{OsbPro}
     \art{Osborne A.R., Provenzale A.}{\PD}{1989}{35}{357}

\bibitem{PSVM}
     \art{Provenzale A., Smith L.A. et al.}{\PD}{1992}{58}{31}

\bibitem{Theiler-etal92}
     \art{Theiler J., Eubank S. et al.}{\PD}{1992}{58}{77}

\bibitem{TheilerPrichard96}
     \art{Theiler J., Prichard D.}{\PD}{1996}{94}{221}

\bibitem{Schreiber:surrogates}
     \art{Schreiber T., Schmitz A.}{\PD}{2000}{142}{346}

\bibitem{Octave}
     \book{Eaton J.W.}{GNU Octave: A High-\hspace{0pt}Level Interactive
     Language for Numerical Computations}{Edition~3 for version
     2.0.13, 1997;\newline \hbox{\tt http://www.octave.org/}}

\bibitem{TISEAN}
     \art{Hegger R., Kantz H., Schreiber T.}{Chaos}{1999}{9}{413}\newline
     \hbox{\tt http://www.mpipks-dresden.mpg.de/\~{}tisean}

\bibitem{Takens85}
     \proc{Takens F.}{Dynamical Systems and Bi\-fur\-cations,
     Groningen, 1984, B.~L.~J.~Braaksma, H.~W.~Broer, and F.~Takens,
     eds., Lecture Notes in Mathematics}{1125}{Berlin, Springer-Verlag}
     {1985}{99}

\bibitem{Theiler88}
     \art{Theiler J.}{\PLA}{1988}{135}{195}

\bibitem{BDSL}
     \textit{Brock W.A., Dechert W.D. et al.}
     A Test for Independence Based on the Correlation Di\-mension.
     University of Wisconsin Press. Ma\-di\-son. 1988.

\bibitem{SS99}
     \art{Schmitz A., Schreiber T.}{Phys. Rev.~E}{1999}{59}{4044}

\bibitem{PK98}
     \art{Potapov A., Kurths J.}{\PD}{1998}{120}{369}

\bibitem{Aglietta}
     \art{Aglietta M., Alessandro B. et al.}{J.~Geophys.
     Research}{1993}{98}{15,241}

\bibitem{Bergamasco92}
     \art{Bergamasco L., Serio M., Osborne A.R.}{J.~Geophys.
     Research}{1992}{97}{17,153}

\bibitem{Bergamasco94}
     \art{Bergamasco L., Serio M., Osborne A.R.}{J.~Geophys.
     Research}{1994}{99}{4235}

\bibitem{Japan2001}
     \proc{Saito S., Chinomi S. et al.}
     {Proc. 27th ICRC}{1}{Hamburg}{2001}{212}

\bibitem{Japan:particles}
     \proc{Chikawa M., Hasebe N. et al.}{Proc. 24th ICRC}{1}{Rome}
     {1995}{277}

\bibitem{Japan:FractalWave}
     \proc{Ohara S., Konishi T. et al.}
     {Proc. 27th ICRC}{1}{Hamburg}{2001}{208}

\end{thebibliography}
\end{document}